\documentclass[sigconf, screen]{acmart}
\AtBeginDocument{
  }

\setcopyright{acmlicensed}
\copyrightyear{2025}
\acmYear{2025}
\setcopyright{cc}
\setcctype{by}
\acmConference[MMAsia '25]{ACM Multimedia Asia}{December 9--12, 2025}{Kuala Lumpur, Malaysia}
\acmBooktitle{ACM Multimedia Asia (MMAsia '25), December 9--12, 2025, Kuala Lumpur, Malaysia}\acmDOI{10.1145/3743093.3770976}
\acmISBN{979-8-4007-2005-5/2025/12}
\usepackage{amsmath}
\usepackage{bm}
\usepackage{gensymb}

\def\btheta{\bm{\theta}}

\def\btheta{\bm{\theta}}
\newcommand{\mytilde}{\raise.17ex\hbox{$\scriptstyle\mathtt{\sim}$}}

\def\aa{\mathbf{a}}

\def\dd{\mathbf{d}}

\def\ff{\mathbf{f}}

\def\ttt{\mathbf{t}}

\def\vv{\mathbf{v}}
\def\ww{\mathbf{w}}

\def\zz{\mathbf{z}}

\def\cC{\mathcal{C}}
\def\dD{\mathcal{D}}

\def\lL{\mathcal{L}}

\def\tT{\mathcal{T}}

\def\Re{\mathbb{R}}

\usepackage{enumitem}
\setlist[itemize]{nosep, leftmargin=*}

\begin{document}

\title{SeeingSounds: Learning Audio-to-Visual Alignment via Text}

\author{Simone Carnemolla}
\email{simone.carnemolla@phd.unict.it}
\orcid{https://orcid.org/0009-0005-8367-3933}
\affiliation{
  \institution{University of Catania}
  \city{Catania}
  \country{Italy}
}

\author{Matteo Pennisi}
\email{matteo.pennisi@unict.it}
\orcid{https://orcid.org/0000-0002-6721-4383}
\affiliation{
  \institution{University of Catania}
  \city{Catania}
  \country{Italy}
}

\author{Chiara Russo}
\email{chiaramaria.russo@studium.unict.it}
\orcid{https://orcid.org/0009-0006-4402-6159}
\affiliation{
  \institution{University of Catania}
  \city{Catania}
  \country{Italy}
}

\author{Simone Palazzo}
\email{simone.palazzo@unict.it}
\orcid{https://orcid.org/0000-0002-2441-0982}
\affiliation{
  \institution{University of Catania}
  \city{Catania}
  \country{Italy}
}

\author{Daniela Giordano}
\email{daniela.giordano@unict.it}
\orcid{https://orcid.org/0000-0001-5135-1351}
\affiliation{
  \institution{University of Catania}
  \city{Catania}
  \country{Italy}
}

\author{Concetto Spampinato}
\email{concetto.spampinato@unict.it}
\orcid{https://orcid.org/0000-0001-6653-2577}
\affiliation{
  \institution{University of Catania}
  \city{Catania}
  \country{Italy}
}

\renewcommand{\shortauthors}{Carnemolla et al.}

\begin{abstract}
We introduce \textbf{SeeingSounds}, a lightweight and modular framework for audio-to-image generation that leverages the interplay between audio, language, and vision—without requiring any paired audio-visual data or training on visual generative models. Rather than treating audio as a substitute for text or relying solely on audio-to-text mappings, our method performs dual alignment: audio is projected into a semantic language space via a frozen language encoder, and, contextually grounded into the visual domain using a vision-language model. This approach, inspired by cognitive neuroscience, reflects the natural cross-modal associations observed in human perception. The model operates on frozen diffusion backbones and trains only lightweight adapters, enabling efficient and scalable learning. Moreover, it supports fine-grained and interpretable control through procedural text prompt generation, where audio transformations (e.g., volume or pitch shifts) translate into descriptive prompts (e.g., “a distant thunder”) that guide visual outputs. Extensive experiments across standard benchmarks confirm that SeeingSounds outperforms existing methods in both zero-shot and supervised settings, establishing a new state of the art in controllable audio-to-visual generation.

\end{abstract}



\maketitle

\section{Introduction}
Generative models conditioned on textual descriptions have led to remarkable progress in image and video synthesis. Models like DALL-E \cite{dall-e} and Stable Diffusion \cite{stable-diffusion} have demonstrated that large-scale training on vision-language data enables highly expressive and coherent generation.
Beyond text conditioning, recent works have explored audio as a complementary modality for visual synthesis, motivated by its natural co-occurrence with visual events and its ability to convey contextual and temporal information.

Early approaches to audio-conditioned image synthesis primarily relied on Generative Adversarial Networks (GANs)~\cite{chen2017deep,hao2018cmcgan,wan2019sound,fanzeres2021sound,soundguided2021,intowild}, showing that sound can guide visual generation but often operating on constrained datasets with strong audio-visual correlations, and focusing more on style manipulation than true synthesis. 
To improve generation quality and scalability, recent works have increasingly adopted diffusion models, leveraging powerful pre-trained text-to-image architectures while introducing audio-to-text mappings to recycle existing language supervision~\cite{gluegen2023,yariv2023adaptation,imagebind2023}. While these approaches are effective, they typically treat language as the sole intermediary—aligning audio to vision indirectly through text. SonicDiffusion~\cite{lee2023robust} takes a step further by injecting audio features directly into the diffusion process via cross-attention, resulting in high-quality audio-guided generation. However, these methods overlook the deeper, complementary relationships between \textit{language, hearing, and vision} - interconnections that are strongly supported by cognitive neuroscience as foundational to human perception~\cite{subramaniam2024visionlanguage,GHAZANFAR2006278,BAUM20121825}.

This evolving line of research coincides with a broader shift in generative modeling: diffusion models now support richer conditioning interfaces, enabled by dedicated encoders for multiple input types. Recent multimodal frameworks like Flux~\cite{flux2024} have extended this paradigm, employing modular adapters and mapping networks to integrate information from diverse sources such as audio, language, and vision.

Building on this recent shift toward multimodal conditioning in generative modeling, we propose \textbf{SeeingSounds}, a lightweight and modular framework for controllable visual generation via \textit{trimodal alignment} between audio, language, and vision. Our approach leverages multi-encoder diffusion architectures and introduces two core text-mediated alignment paths: (i) an \textit{audio-to-text} mapping through a frozen language model, and (ii) a \textit{text-to-vision} grounding via CLIP. This design enables our model to generate images directly from audio, without relying on paired audio-visual data during training. Crucially, conversely to existing diffusion-based approaches, we do not train or fine-tune the diffusion model, only lightweight adapters on top of frozen backbones are optimized.

A key strength of our framework is its controllability, enabled by the intermediate text-based alignment. Audio modifications are translated into textual prompts—e.g., lowering the volume of a train sound yields \textit{“the sound of a distant train”}—which guide generation in a semantically consistent and interpretable way. This allows for fine-grained control without modifying or fine-tuning the generative model.

We evaluate it across a wide set of benchmarks — VEGAS~\cite{vegas}, VGGSound~\cite{vggsound}, Landscape~\cite{landscape} combined to  Into the Wild~\cite{intowild} and RAVDESS~\cite{ravdess} — spanning diverse acoustic environments, material interactions, natural scenes, and emotional expressions. 
Our results show state-of-the-art performance in audio-to-scene generation, both qualitatively and quantitatively, confirming the effectiveness of our simple yet effective \textit{language-vision-audio} alignment strategy.\\
To summarize, the contributions of this work are:
\begin{itemize}[itemsep=1pt, parsep=1pt, topsep=4pt]
\item A lightweight and modular framework for visual generation from audio, trained without the need for any paired audio-visual data.
\item An alignment strategy that bridges audio, language, and vision through semantically grounded connections.
\item An interpretable and controllable generation pipeline that allows procedural audio modifications to be reflected in the visual output via text-mediated prompt manipulation;
\item A thorough empirical evaluation demonstrating strong generalization and significantly outperforming prior methods on zero-shot audio-to-image benchmarks.
\end{itemize}

\section{Related work}
Audio-driven visual synthesis spans image and video generation. Early work mainly used VAEs/GANs to bridge audio–image semantics~\cite{manifoldlearning2022,chen2017deep, fanzeres2021sound, hao2018cmcgan, intowild, wan2019sound,soundguided2021}. Notably, SGSIM~\cite{soundguided2021} and Sound2Scene~\cite{sound2scene2023} injected CLIP/ResNet audio embeddings into pretrained GANs (e.g., StyleGAN/BigGAN) with contrastive losses, but struggled on in-the-wild data, with low semantic fidelity and weak generalization. Robust-SGSIM~\cite{lee2023robust} added KL-regularization between audio–image and text–image cosine-similarity distributions, among the first to combine audio–text–image. AVStyle~\cite{intowild} used audiovisual adversarial training and patch-wise contrastive learning for stylization. Despite these advances, diffusion models have surpassed GANs in quality, fidelity, and scalability. Diffusion-era methods condition generation on audio via learned multimodal spaces. GlueGen~\cite{gluegen2023} fuses XLM-R (text) and AudioCLIP~\cite{audioclip} features, aligned in CLIP space for latent diffusion. AudioToken~\cite{yariv2023adaptation} injects audio directly into CLIP’s text-token space. These rely on indirect audio–image alignment. ImageBind~\cite{imagebind2023} learns a joint space for audio–image–text (and more) via InfoNCE, but targets retrieval/representation rather than controllable generation. SonicDiffusion~\cite{sonicdiffusion2024} adapts Stable Diffusion with audio-aware cross-attention for high-quality audio-guided synthesis/editing, yet mediates audio through text, without explicit trimodal alignment.

We introduce a lightweight trimodal framework that semantically aligns audio, language, and vision. Using frozen encoders with small adapters trained on disjoint audio–text and text–image corpora, we avoid any paired audio–visual data. Our approach offers fine-grained, interpretable control via procedural prompt manipulation: audio changes (e.g., pitch/volume) map to textual prompts (e.g., “a distant thunder”) that steer generation—without tuning the generative model.

\section{Method}
\begin{figure*}[ht]
    \centering
    \includegraphics[width=0.60\textwidth]{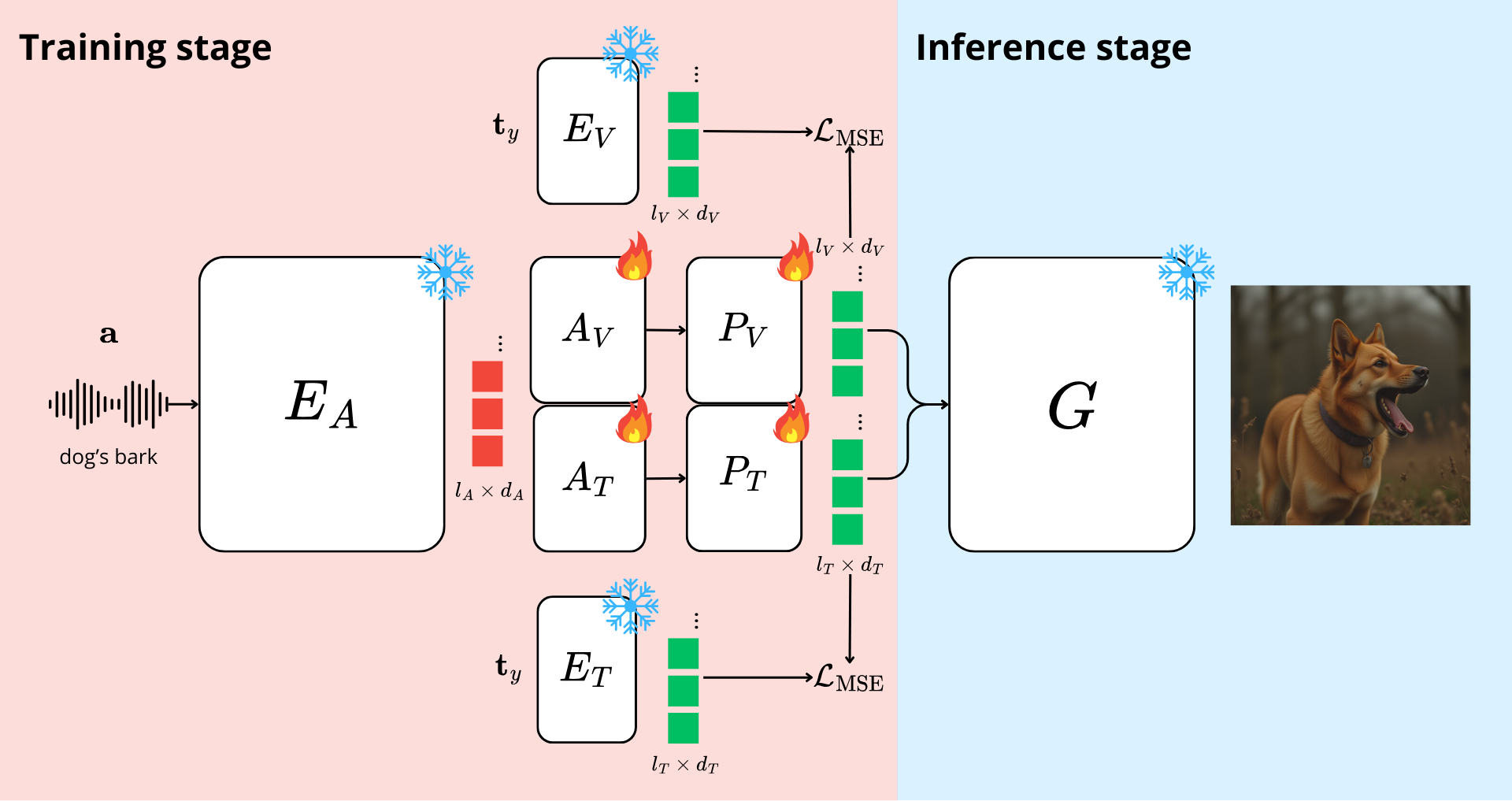}
    \caption{Overview of audio-text and visual modal alignment in Seeing Sounds. During training (left), audio features $E_A$ are aligned to both language and vision-language spaces via lightweight adapters ($A_T$ and $A_V$) and projected (through $P_T$ and $P_V$) using MSE losses against frozen text ($E_T$)and vision-text ($E_V$) encoders. At inference (right), the audio-text-vision-aligned projection conditions a frozen diffusion model for audio-driven image generation, enabling controllable and paired-free synthesis.}
    \label{fig:Method}
\end{figure*}

We propose an alignment framework for audio-conditioned visual generation that leverages language as an intermediary to connect the auditory and visual domains. Central to our approach is our alignment strategy: one branch of the architecture aligns audio features to textual semantics, while the other aligns audio features to a vision-aware latent space via language. This enables the model to benefit from the compositional structure of natural language and the grounding power of large-scale vision-language models, without requiring the need of visual data at training time. Furthermore, unlike existing and recent methodologies \cite{sonicdiffusion2024, audiotoken2023}, our pure alignment-based design does not involve the visual generation model in any way, allowing for a more efficient training paradigm.

Let \( \dD  = \left\{(\aa_1, y_1), (\aa_2, y_2), \ldots, (\aa_n, y_n)\right\} \) be an audio dataset, where each audio sample \( \aa_i \) is associated with a class label \( y_i \in \cC \), \( \left| \cC \right| = k \). For each class in \( \cC \), we also assume the availability of a corresponding textual description, building up a set \( \tT = \{\ttt_1, \ttt_2, \dots, \ttt_k\} \) of ``class-level captions''. This requirement aligns with similar approaches in the literature~\cite{sonicdiffusion2024}, which utilize image data with image-level captions obtained via pre-trained captioning models. Our approach relaxes this requirement by needing only a class-level caption for each class obtained by prompting a large language model (system prompts are reported in supplementary materials).

Given an audio sample \( (\aa, y) \), we employ a transformer-based audio encoder \( E_A \) as a feature extractor, to derive a latent representation $\ff$ in the form of a sequence of tokens:
\begin{equation}
\ff = E_A(\aa) = \left[ \ff_1, \ff_2, \ldots, \ff_m \right], 
\end{equation}
where each $\ff_i \in \Re^{d_A}$ and $d_A$ is the dimensionality of each audio token. Our goal is to align this audio representation with the representations generated by a text encoder \( E_T \) and a vision-language encoder \( E_V \), both conditioned by the class captions in \( \tT \).

The text encoder \( E_T \) is a pure textual model that, given sample $\left( \aa, y \right)$, processes a tokenized version of the class-level caption \( \ttt_y \), transforming it into a sequence of output tokens $\zz_T = E_T(\ttt_y) = \left[ \zz_{T,1}, \zz_{T,2}, \dots, \zz_{T,l} \right]$, with each $\zz_{T,i} \in \Re^{d_T}$, where \( d_T \) is the dimensionality of the textual representation. In contrast, the vision-language encoder \( E_V \), while being similarly conditioned by the caption $\ttt_y$, is designed to produce a vision-aware representation of the input text, as done by CLIP \cite{radford2021learning} and similar models. The output of the vision-language encoder $E_V$ is a vector $\zz_V = E_V(\ttt_y) \in \Re^{d_V}$, with $d_V$ being the dimensionality of the embedding space.

To facilitate the alignment of audio features \( \ff \) with \( \zz_T \) and \( \zz_V \), we introduce adapter modules \( A_T \) and \( A_V \) to project audio feature tokens onto the same dimensionality as the target representations. Each module is implemented as a multi-layer perceptron, producing the corresponding adapted features $\ww_T$ and $\ww_V$:
\begin{equation}
\ww_T = A_T(\ff) = \left[ \ww_{T,1}, \ww_{T,2}, \dots, \ww_{T,m} \right],
\end{equation}
\begin{equation}
\ww_V = A_V(f) = \left[ \ww_{V,1}, \ww_{V,2}, \dots, \ww_{V,m} \right],
\end{equation}
with $\ww_{T,i} \in \Re^{d_T}$ and $\ww_{V,i} \in \mathbb{R}^{d_V}$.

The final step before feature alignment consists in matching the lengths of the sequences in $\ww_T$ and $\ww_V$ with those of, respectively, $\zz_T$ and $\zz_V$. Indeed, $\ww_T$ and $\ww_V$ come directly from the encoded input sequence $\ff$, which contains $m$ tokens; instead, the number of tokens in $\zz_T$ depends on the caption $\ttt_y$ (which has $l$ tokens), while $\zz_V$ is a single vector. To solve this discrepancy, we employ attention pooling~\cite{attentive-pooling}, which allows for a flexible mapping between the different lengths of the sequences. Attention pooling projects a sequence of \( p \) tokens to \( q \) tokens, utilizing \( q \) learnable query tokens and performing self-attention over the entire set of tokens to learn the mapping from the original \( p \) tokens. This method is preferred over alternative pooling strategies (e.g., token averaging or concatenation) due to its ability to capture complex relationships between tokens, as well as its natural application to token-structured data.

Let \( P_{p,q,\btheta} \) denote the attention pooling operation, which transforms a sequence of \( p \) tokens into \( q \) tokens using parameters \( \btheta \). The pooled representations are computed as follows:
\begin{equation}
\hat{\zz}_T = P_{m,l,\btheta_T}(\ww_T) = P_T(\ww_T) = [\hat{\zz}_{T,1}, \ldots, \hat{\zz}_{T,l}] ,
\end{equation}
\begin{equation}
\hat{\zz}_V = P_{m,1,\btheta_V}(\ww_V) = P_V(\ww_V) \in \Re^{d_V},
\end{equation}
where $\btheta_T$ and $\btheta_V$ are the parameters of the two pooling layers (i.e., the learned query tokens and the cross-attention projection matrices), and each $\hat{\zz}_{T,j} \in \Re^{d_T}$.

The resulting representations \( \hat{\zz}_T \) and \( \hat{\zz}_V \) are subsequently fed into a visual generator \( G \) to produce the output image $\vv = G(\hat{\zz}_T, \hat{\zz}_V)$.

During the training phase, we optimize the parameters of the adapter modules and attention pooling layers, while keeping the encoders \( E_A \), \( E_T \), \( E_V \), and the generator \( G \) frozen. 
Let \( \hat{z}_T(\cdot) \) and $\hat{z}_V(\cdot)$ denote the process of computing \( \hat{\zz}_T \) and \( \hat{\zz}_V \) from $\aa$; similarly, let \( {z}_T(\cdot) \) and ${z}_V(\cdot)$ compute the target representations \( {\zz}_T \) and \( {\zz}_V \) from $y$ (and hence $\ttt_y$).
The alignment of the learned representations is achieved by minimizing the mean square error (MSE) loss between the generated and target representations:
\begin{equation}
\begin{aligned}
\lL_\text{MSE} & = \frac{1}{n} \sum_{(\aa,y)} \left[ \| \hat{z}_T\left( \aa \right) - z_T \left( \ttt_y \right) \|^2 + \| \hat{z}_V \left( \aa \right) - z_V \left( \ttt_y \right) \|^2  \right] \\
                 & = \frac{1}{n} \sum_{(\aa,y)} \lL \left( \aa, \ttt_y \right),
\end{aligned}
\label{eq:loss}
\end{equation}
with $\lL$ representing, for brevity, the loss contribution of a single sample.

This training strategy enables an effective and efficient model training process, maintaining a limited number of learnable parameters and circumventing the need for backpropagation through the diffusion model, which is typically computationally intensive.

\subsection{Enhancing controllability}
\label{sec:controllability}
Since the generation process is mediated by textual semantics, without any requirement for training images, we can guide the audio-to-visual generation process by manipulating the input audio signal and the associated class-level captions. This allows us to effectively generate additional ``pseudo-classes'', extending the range of audio variability that our model can render into visual scenes. For instance, consider an audio recording of a train: the associated class caption might be ``a train is passing by''. By modifying the audio signal --- e.g., attenuating the volume or adding reverberation --- we can correspondingly change the descriptive text to ``a distant train'' or ``a train echoing in a tunnel''. Importantly, this whole procedure can be carried out programmatically, by predefining the set of augmentation effects, and computing the associated captions through a large language model (LLM).
An added benefit of this procedure is that it also enables the capability of modeling \emph{mixed} audio signals by combining the respective descriptions, unlike other approaches that require non-trivial and unrealistic image-level superimposition.

Formally, let $(\aa, y)$ be a sample from the audio dataset, and $\ttt_y$ the corresponding class-level caption. Let $\left\{ \left(T_1, \dd_1 \right), \dots, \left( T_{h}, \dd_{h} \right) \right\}$ be a set of audio transformations: to each transformation function $T_j$, a textual \emph{transformation description} $\dd_j$ is associated, which describes the effect of applying $T_j$ to an audio sample $\aa$.
Then, let $f$ be a text transformation function that, given a class-level caption $\ttt_y$ and a transformation description $\dd_j$, stochastically produces an updated version of $\ttt_y$, in light of applying the transformation described by $\dd_j$.

Additionally, let $(\aa_i, u)$ and $(\aa_j,v)$ be two samples from the dataset, and $g$ be another text transformation function, which acts on the class-level captions $\ttt_u$ and $\ttt_v$ to produce a description $\ttt_{u+v}$ of a scene featuring the concepts described by the captions: our assumption is that $\ttt_{u+v}$ is a plausible description of the audio signal obtained as $\aa_i + \aa_j$.

The generated semantically-consistent pairs --- procedurally modified audio and edited captions --- are used to supervise the training process, encouraging the model to capture nuanced audio-text relationships. The alignment objective in Eq.~\ref{eq:loss} is updated to include this extended set:
\begin{equation}
\begin{aligned}
\lL_\text{ext} & = \frac{1}{n} \sum_{(\aa,y)} \lL \left( \aa, \ttt_y \right) + \\
& +\frac{1}{nh} \sum_{(\aa,y)} \sum_{j=1}^h \lL\left( T_j\left( \aa  \right) , f\left( \ttt_y, \dd_j \right)  \right) + \\ 
& + \frac{1}{n(n-1)} \sum_{\substack{(\aa_i,u),\\(\aa_j,v)}} \lL \left( \aa_i + \aa_j, g\left( \ttt_u, \ttt_v \right) \right) .
\end{aligned}
\end{equation}

In practice, we manually provide the transformation descriptions, while $f$ and $g$ are implemented by prompt-engineering an LLM. Details are provided in the supplementary materials.

This formulation enables the model to learn a disentangled representation space, where variations in the audio domain are reflected in semantically meaningful changes in the generated visuals, improving generalization capabilities even with synthetic supervision.

\section{Experimental Results}
\subsection{Training/evaluation setup}
\label{training_evaluation_setup}

We use AST \cite{gong2021ast}, pretrained on AudioSet, as audio encoder and Flux diffusion \cite{flux2024} with a pretrained LoRA adapter for image generation. Text prompts are embedded via CLIP \cite{radford2021learning} and T5 \cite{2020t5}. During training, image generation is disabled to speed up convergence, and model selection is based on validation loss. Prompts for each dataset class are generated with GPT-4o, using standardized system prompts (details in supplementary), and follow best practices in prompt engineering \cite{wei2022chain, brown2020language, amatriain2024prompt, wenjuan-etal-2024-prompt, ramesh2021zero}. Due to different encoder dynamics, we use separate AdamW optimizers: T5 parameters use a $1 \times 10^{-5}$ learning rate; CLIP-aligned ones use $1 \times 10^{-7}$. All experiments run on a single NVIDIA H100 GPU. We strictly follow training and evaluation protocols from prior work and use reported competitor results. 

For VGGSound \cite{vggsound} and VEGAS \cite{vegas}, we adopt the Sound2Scene protocol \cite{sound2scene2023}, using the 50-class VGGSound subset with a 90/10 validation split, and both full and 5-class versions of VEGAS (800/50 samples per class for train/test). For RAVDESS \cite{ravdess} and Into the Wild + Landscape \cite{intowild, landscape}, we apply the training procedures in \cite{sonicdiffusion2024}, using 90/10 splits. Early stopping is applied across all benchmarks. We also test zero-shot performance on ESC-50 \cite{esc-50}, generating one image per sample (2,000 total) without training. Competitors are evaluated using their published results, except: 1) AudioToken was retrained on VGGSound-50 for comparability with \cite{sound2scene2023}; 2) ImageBind was evaluated using its pretrained weights and Stable Diffusion UnCLIP, as in \cite{audiotoken2023, sonicdiffusion2024}. For ESC-50, we retain Sound2Scene and AudioToken original weights. Inference uses mixed-precision for efficiency. We fix diffusion steps to 10, with 3.10s average generation time per image.

While \cite{sonicdiffusion2024} include the Greatest Hits dataset~\cite{greatesthits}, we exclude it from quantitative evaluation due to insufficient protocol details, providing only qualitative results for illustration in the supplementary.

For controllability tests, we scale input audio using a gain factor $\alpha \in [0.1, 0.5]$, with corresponding labels: low-volume'' ($\alpha < 0.2$), medium-volume'' ($0.2 < \alpha < 0.5$), and ``high-volume'' (otherwise). In mixed controllability, input audio combines samples from two classes, and the caption reflects both, enabling evaluation on ambiguous content.

Additional dataset and metric details can be found in the supplementary and at the following link: \href{https://seeingsounds.netlify.app/}{SeeingSounds supplementary}.
 
\begin{figure}[h!]
    \centering
    \includegraphics[width=0.8\linewidth]{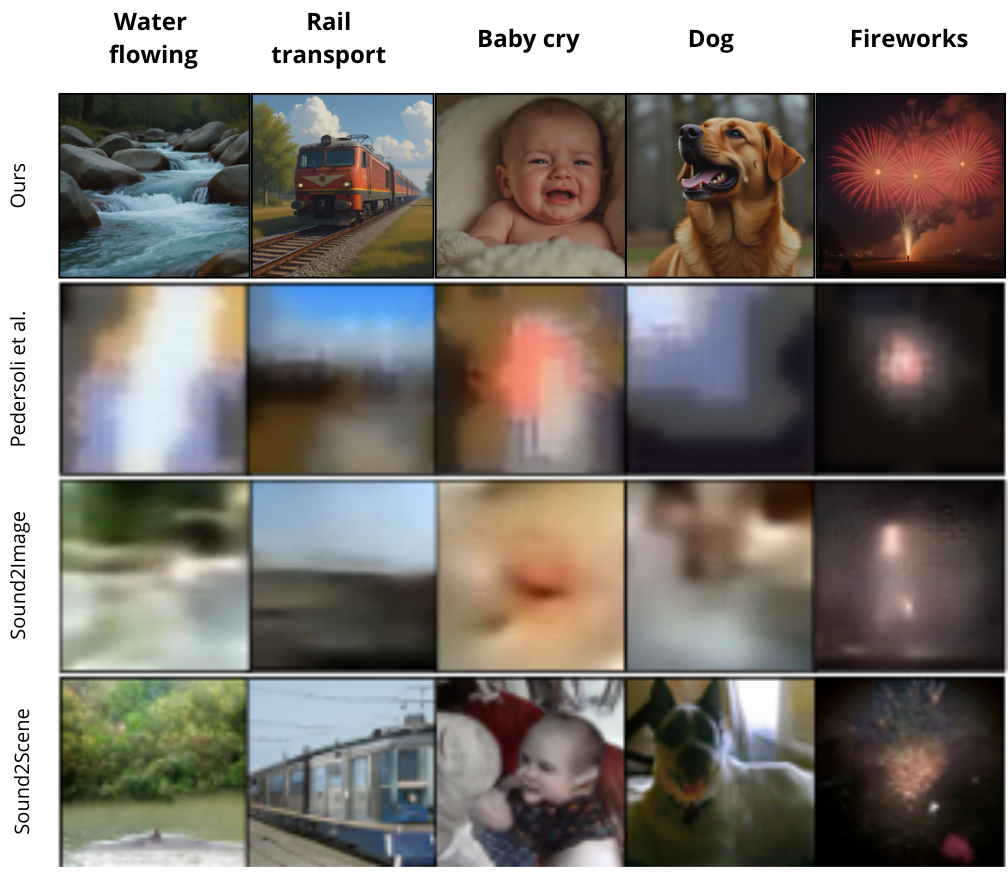}
    \caption{\textbf{Qualitative results on VEGAS (5-class subset).} Our model yields semantically accurate generations aligned with the input audio category. Images adapted from~\cite{sound2scene2023} to include our results.}
    \label{fig:examples-vegas}
\end{figure}

\subsection{Results}

\textbf{Comparison to state of the art methods.} We start our evaluation by comparing the performance of our method with several state-of-the-art methods, including GAN-based techniques (e.g., SGSIM~\cite{soundguided2021}, Sound2Scene~\cite{sound2scene2023}), embedding-based methods (e.g., AudioToken~\cite{audiotoken2023}, ImageBind~\cite{imagebind2023}), and recent diffusion-based approaches (e.g., GlueGen~\cite{gluegen2023}, SonicDiffusion~\cite{sonicdiffusion2024}, TempoTokens~\cite{audiovideogen2023}, CoDi~\cite{composablediffusion2023}). Evaluation is conducted using two key metrics: AIC (Audio-Image Content) and AIS (Audio-Image Similarity), where higher scores indicate stronger audio-visual alignment. As mentioned in Section \ref{training_evaluation_setup}, we rely exclusively on results reported in the original publications of each method. Consequently, not all baselines are evaluated on every dataset due to differences in experimental setups, training requirements, or dataset compatibility across methods. Table~\ref{tab:vggsound_vegas} reports performance on the VGGSound (50-class subset) and the VEGAS dataset, including its smaller 5-class version proposed in~\cite{sound2scene2023}. 
Our model achieves a substantial improvement over all baselines, with about 25 AIC and 14 R@5 gain over the best previous methods on VGGSound, and a large margin of AIC and R@5 gain also on both versions of VEGAS. 
\begin{table}[]
\caption{\textbf{Comparison with state-of-the-art models on VGGSound and VEGAS.} We report AIC and Recall@5 (R@5) on the 50-class subset of VGGSound and the full and 5-class VEGAS datasets respectively.}
\label{tab:vggsound_vegas}
\resizebox{0.9\columnwidth}{!}{
\begin{tabular}{@{}lcc|cc|c@{}}
\toprule
\multicolumn{1}{c}{\textbf{}} & \multicolumn{2}{c|}{\textbf{VGGSound}} & \multicolumn{2}{c|}{\textbf{VEGAS}} & \textbf{VEGAS5} \\ \midrule
\multicolumn{1}{c}{}                 & \textbf{AIC}   & \textbf{R@5}   & \textbf{AIC}   & \textbf{R@5}   & \textbf{AIC}   \\
\midrule
{Pedersoli et al.} \cite{manifoldlearning2022} & -              & -              & -              & -              & 23.10          
\\
S2I    \cite{fanzeres2021sound}                              & -              & -              & -              & -              & 39.19          \\
ICGAN    \cite{instancegan2021}                            & 30.06          & 62.59          & 46.60          & 82.48          & -              \\
Sound2Scene    \cite{sound2scene2023}                      & 40.71          & 77.36          & 57.44          & 84.08          & 77.58          \\
ImageBind    \cite{imagebind2023}                      & 40.62              &     74.72       & -              & -              & -              \\
AudioToken    \cite{audiotoken2023}                       &   41.30             &  77.42          & -              & -              & -              \\
\textbf{Ours}                        & \textbf{66.64} & \textbf{91.60} & \textbf{90.81} & \textbf{96.28} & \textbf{94.30} \\ \bottomrule
\end{tabular}
}
\end{table} 
Table~\ref{tab:sonic_datasets} presents a broader comparison on RAVDESS and Landscape + Into the Wild. Our method consistently outperforms prior approaches, achieving the highest AIC scores on both RAVDESS (35.39) and Landscape + Into the Wild (79.13), with substantial margins over baselines such as SonicDiffusion and GlueGen. While our AIS score is lower than ImageBind's one, it is worth noting that ImageBind is a foundation model specifically trained on large-scale multimodal data to perform cross-modal alignment. In contrast, our method achieves stronger semantic fidelity—as reflected in AIC—using lightweight alignment modules and without any end-to-end training of the diffusion model.

\begin{table}[]
\caption{\textbf{Cross-dataset comparison on RAVDESS, Landscape + Into the Wild, and Greatest Hits.} Our model achieves the highest AIC across all benchmarks, despite using no paired audio-visual training data or diffusion model fine-tuning.}
\label{tab:sonic_datasets}
\resizebox{\columnwidth}{!}{
\begin{tabular}{@{}lllcc@{}}
\toprule
\multicolumn{1}{c}{\textbf{}} &
  \multicolumn{2}{c}{\textbf{RAVDESS}} &
  \multicolumn{2}{c}{\textbf{Into the wild + Landscape}} \\ \midrule
&
  \textbf{AIS} & \multicolumn{1}{c|}{\textbf{AIC}} &
  \textbf{AIS} &
  \textbf{AIC} \\
  \midrule
SGSIM \cite{soundguided2021}          & 50.53                 & \multicolumn{1}{l|}{12.57}          & 72.24 & 21.66 \\
Sound2Scene \cite{sound2scene2023}   & 50.90                 & \multicolumn{1}{c|}{11.08}          & 74.66 & 36.72 \\
GlueGen  \cite{gluegen2023}     & 50.52                 & \multicolumn{1}{l|}{22.52}          & 66.32 & 46.18 \\
ImageBind \cite{imagebind2023}     & \textbf{58.02}        & \multicolumn{1}{l|}{21.31}          & 72.09 & 46.18 \\
CoDi  \cite{composablediffusion2023}         & 52.92                 & \multicolumn{1}{l|}{13.96}          & 75.78 & 46.00 \\
AudioToken \cite{audiotoken2023}   & 50.09                 & \multicolumn{1}{c|}{18.21}          & 69.83 & 36.18 \\
SonicDiffusion \cite{sonicdiffusion2024} & 53.09                 & \multicolumn{1}{l|}{23.16}          & 74.46 & 28.51 \\
TempoTokens \cite{audiovideogen2023}   & \multicolumn{1}{c}{-} & \multicolumn{1}{c|}{-}              & 73.90 & 54.36 \\
\textbf{Ours}  & 51.41                 & \multicolumn{1}{c|}{\textbf{35.39}} & \textbf{76.16} & \textbf{79.13} \\ \bottomrule
\end{tabular}
}
\end{table}

As shown in Tables~\ref{tab:vggsound_vegas} and~\ref{tab:sonic_datasets}, our model achieves consistent and substantial improvements across all datasets through a lightweight training pipeline that does not rely on visual inputs or paired audio-image data. These results cumulatively demonstrate the power and flexibility of our trimodal alignment strategy, which enables visually coherent and semantically grounded generation from sound without training large generative backbones.

We further assess the effectiveness of our approach through qualitative comparisons. Figure~\ref{fig:examples-vegas} shows generations from the VEGAS 5-class subset, where our method produces visually coherent and semantically appropriate scenes across diverse sound categories, outperforming previous methods in clarity, object fidelity, and spatial composition. While part of this improvement may stem from leveraging a diffusion-based architecture over less expressive GANs, 
supplementaries extend the comparison to more challenging datasets—including RAVDESS, Landscape + Into the Wild, and Greatest Hits—where we benchmark against recent diffusion-based methods such as SonicDiffusion, AudioToken, and GlueGen. Our model consistently produces high-quality, context-aware visuals that reflect subtle audio cues, such as emotional intonation or material properties. These visual comparisons highlight the expressive capacity and generalization ability of our controllable, audio-conditioned generation pipeline.

\noindent\textbf{Zero-shot performance.} To evaluate the generalization capacity of our method beyond the training distribution, we conduct a zero-shot image generation experiment on the ESC-50 dataset, which comprises environmental sound clips from diverse categories such as animal calls, machinery, weather, and human vocalizations. Crucially, this dataset is never seen during training, making it ideal for testing out-of-distribution performance. We compare our approach against existing methods that release pretrained weights suitable for inference-time use, ensuring a fair and reproducible comparison without retraining. This constraint limits the pool of competitors but avoids discrepancies caused by inconsistent training pipelines. As shown in Table~\ref{tab:zero-shot}, our method achieves a substantial improvement over prior work—more than doubling the AIC and R@5 scores of the strongest baseline (AudioToken~\cite{audiotoken2023}). 

\begin{table}[]
\caption{\textbf{Zero-shot evaluation on the ESC-50 dataset.} Our model achieves significantly higher AIC and Recall@5 scores than prior works, demonstrating superior generalization to unseen audio domains.}
\label{tab:zero-shot}
\resizebox{0.5\columnwidth}{!}{

\begin{tabular}{@{}llcc@{}}
\toprule
\multicolumn{1}{c}{\textbf{}} & \multicolumn{2}{c}{ESC50}\\ \midrule
\multicolumn{1}{c}{}                 & \textbf{AIC}   & \textbf{R@5}   \\
Sound2Scene  \cite{sound2scene2023}                                  & 4.10           & 16.40          \\
AudioToken    \cite{audiotoken2023}               & 9.00           & 22.00          \\
\textbf{Ours}                                     & \textbf{22.60} & \textbf{41.35} \\ \bottomrule
\end{tabular}
}
\end{table}
\begin{figure}[ht]
    \centering
    \includegraphics[width=0.8\linewidth]{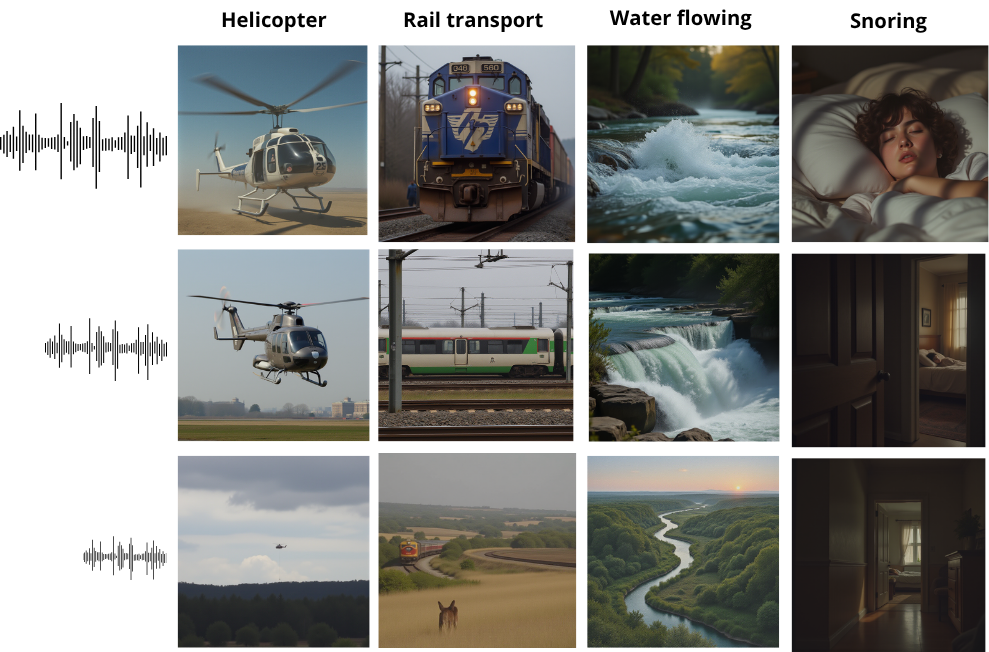}
    \caption{Volume controllability: controllable generation based on audio intensity. Lower volume yields smaller or distant visual representations.}
    \label{fig:examples-volume}
\end{figure}

\begin{figure}[ht]
    \centering
    \includegraphics[width=0.8\linewidth]{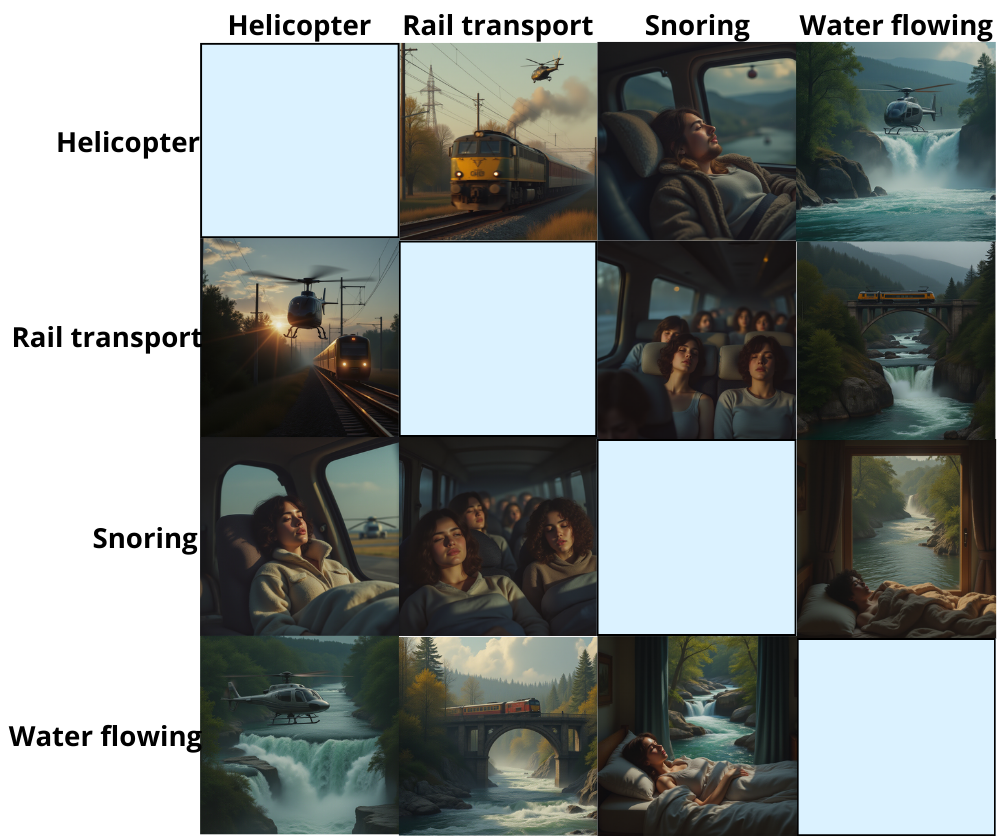}
    \caption{Mixed-class controllability. Each cell visualizes a scene combining the row and column audio classes (e.g. helicopter flying + rail transport).}
    \label{fig:examples-mixed}
\end{figure}

\noindent\textbf{Controllability.} To assess the effectiveness of our controllable generation mechanism described in Sect.~\ref{sec:controllability}, we qualitatively explore how structured manipulations in the audio domain—either through signal transformations or compositional mixing—result in corresponding and semantically aligned changes in the generated visual output. Fig.~\ref{fig:examples-volume} shows the controllability of our model with respect to audio intensity. Specifically, we apply a simple volume attenuation to an audio recording (e.g., of a train or a helicopter) and modify its associated text prompt accordingly (e.g., from ``a train is passing'' to ``a distant train''). As illustrated in the generated samples, our method is capable of reflecting such subtle, continuous auditory variations into the visual modality, producing semantically faithful differences in scale, perspective, or contextual framing. For instance, quieter inputs often yield smaller or more distant visual objects, consistent with human perceptual expectations.
In Figure~\ref{fig:examples-mixed}, we demonstrate the model's ability to synthesize scenes from \textit{compositional audio inputs}. Each cell in the grid corresponds to a generated image conditioned on the mix of two distinct audio sources (e.g., ``train'' and ``helicopter'') and a compositional prompt automatically derived via LLM-assisted caption fusion (e.g., “a distant train and a hovering helicopter”). The model exhibits the capacity to visually represent both acoustic sources in a coherent scene, often arranging them spatially or contextually in a way that matches the semantic structure of the textual prompt. This compositional controllability is particularly challenging for conventional approaches, which often require unrealistic pixel-level superimpositions or explicit supervision.

\begin{table}[]
\caption{Ablation study on the VEGAS 5-class subset. We assess the impact of removing attention pooling and isolating each alignment branch. Results confirm the benefit of both attention mechanisms and the full tri-modal alignment.}
\label{tab:ablation}
\resizebox{0.7\columnwidth}{!}{
\begin{tabular}{@{}l|c@{}}
\toprule
\textbf{Model Variant}                & \textbf{AIC} \\ \midrule
w/o attention pooling (adaptive + mean)   & 54.72        \\
Only CLIP alignment                  & 51.10        \\
Only T5 alignment                    & 91.00        \\
\textbf{Ours (full: T5 + CLIP + attn pooling)} & \textbf{94.30} \\ \bottomrule
\end{tabular}
}
\end{table}

\noindent\textbf{Ablation.} We finally perform a focused ablation study on the 5-class VEGAS subset to validate our architectural choices. As shown in Table~\ref{tab:ablation}, we first examine the role of the attention pooling modules by replacing them with simpler alternatives: adaptive pooling for the T5 branch and mean pooling for the CLIP branch of FLUX. This leads to a notable drop in AIC, confirming that our cross-modal attention pooling layers play a critical role in semantic alignment.
When isolating each alignment branch, we observe that using only T5-based alignment outperforms the CLIP-only variant, indicating that language supervision provides stronger semantic grounding when used alone. Nonetheless, the full model—combining both alignments with attention pooling—achieves the best performance, demonstrating the complementary nature of linguistic and visual semantics in guiding audio-to-visual generation.
Supplementary materials report a first extension on video-generation, multiple audio-image and audio-video samples including qualitatively results on all the employed benchmarks.

\section{Conclusion}
In this work, we introduced \textbf{SeeingSounds}, a lightweight and scalable framework for audio-conditioned image 
generation that unifies audio, language, and vision through a tri-modal alignment strategy. Departing from prior methods that rely on paired audio-visual data or require training large generative backbones, our approach leverages frozen pre-trained encoders with minimal trainable components to align audio to text via a semantic language model and text to vision through CLIP. This indirect yet effective grounding strategy eliminates the need for costly paired data and supports zero-shot generation. Extensive evaluations across multiple benchmarks demonstrate that our method consistently outperforms existing approaches, particularly in challenging low-resource or disjoint-modality settings, while maintaining strong semantic fidelity and generalization. Moreover, the introduction of procedurally generated captions tied to audio augmentations enables fine-grained controllability over the generation process. Our model can respond meaningfully to subtle changes in the input sound or composite multiple audio sources into coherent visual scenes.

Overall, SeeingSounds underscores the effectiveness of language as a bridge for multimodal grounding and open new directions for scalable, controllable, and cognitively inspired generative models.

\begin{acks}
Simone Carnemolla, Matteo Pennisi, Simone Palazzo, Daniela Giordano, and Concetto Spampinato have been supported by the European Union – Next Generation EU, Mission 4 Component 2 Line 1.3, through the PNRR MUR project PE0000013 – FAIR “Future Artificial Intelligence Research” (CUP E63C22001940006).
\end{acks}

\bibliography{biblio}
\bibliographystyle{ACM-Reference-Format}
\end{document}